\documentclass[nofootinbib,twocolumn]{rich2004}

\usepackage{epsfig}
\usepackage{graphicx}
\usepackage{dcolumn}

\usepackage{amssymb}

\begin{document}

\begin{frontmatter}



\title{Performance of the LiF-TEA Ring
Imaging Cherenkov Detector at CLEO}


\author{R. Sia\corauthref{cor1}}
\corauth[cor1]{ For the CLEO RICH group: M. Artuso, R. Ayad, K.
Bukin, A. Efimov, C. Boulahouache, E. Dambasuren, S. Kopp, R.
Mountain, G. Majumder, S. Schuh, T. Skwarnicki, S. Stone, G.
Viehhauser, J.C. Wang, T. Coan, V. Fadeyev, I. Volobouev, J. Ye,
S. Anderson, Y. Kubota, A. Smith.}

\address{Department of Physics, Syracuse University, Syracuse, New York 13244\\
E-mail: rsia@phy.syr.edu}

\begin{abstract}
We describe the particle identification capability of the CLEO
RICH system. This system consists of a 1 cm thick LiF radiator
coupled to a photon detector that uses wire proportional chambers
filled with a mixture of CH$_4$ and TEA. We discuss the yield of
photoelectrons observed per ring and the angular resolution. We
show the efficiencies achieved for particle identification and the
associated fake rates from data collected with both CLEO III and
CLEO-c detectors. Finally we show examples of the particle
separation ability which is excellent for both CLEO III and CLEO-c
data.
\end{abstract}

\end{frontmatter}

\section{Introduction}
In 1999 a particle identification system based on Ring Imaging
Cherenkov detector (RICH) technology, a new vertex detector and a
new wire drift chamber were added to the CLEO detector to probe
physics of the decays of $b$ and $c$ quarks, $\tau$ leptons and
$\Upsilon$ mesons produced near $10~\rm GeV$ in $e^+e^-$
collisions \cite{Artu98}. With RICH, the goal was to achieve a
$\pi/K$ separation greater than $4\sigma$ up to $2.65~\rm GeV/c$,
the mean momentum for pions from $B \to \pi\pi$ decays. At this
momentum, the Cherenkov angle between $\pi$ and $K$ in LiF differs
by $14.3~\rm mrad$. We expect to have in a $1~\rm cm$ thick sample
at this momentum about 12 detected photons with a resolution per
photon of $14~\rm mrad$ yielding to an angular resolution per
track of $4~\rm mrad$. This resolution is enough for more than
$3\sigma$ separation in addition to the $2\sigma$ separation
provided by dE/dx information in the drift chamber for momenta
higher than $2.2~\rm GeV/c$. In this paper, we compare these
expectations with the physics performance of the detector that has
been used for more than 4 years in CLEO III and recently in CLEO-c
at lower center-of-mass energies.

\section{Detector Description}
\begin{figure} [hbt]
\hspace{0cm} \centerline{\epsfxsize
2.4in\epsffile{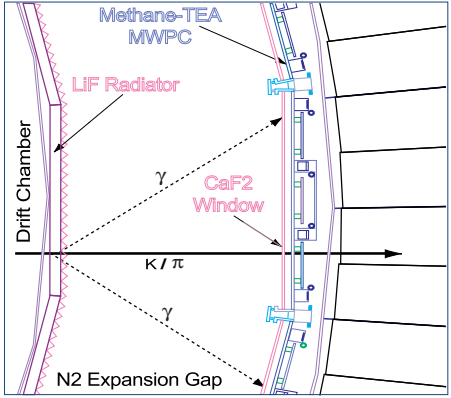}}
\caption{\label{rich-detector} Sketch of the Cherenkov photon
process in the CLEO RICH Detector} \vspace{0.25cm}\end{figure}

Cherenkov photons are produced in a shell of $1~\rm cm$ thick LiF
crystal radiators which, up to a $22^o$ angle from the center of the
solenoid, are sawtooth radiators \cite{efimov} and the rest planar.
The sawtooth radiators are used to prevent total internal reflection
of Cherenkov photons where charged particles cross the detector at
near normal incidence. The photons then enter a $15.6~\rm cm$ thick
expansion volume filled with pure N$_2$ gas, as shown in
Fig.~\ref{rich-detector}, and get detected in multiwire proportional
chambers.
 The
chambers have CaF$_2$ windows and are filled with TEA dispersed in
methane that has the ability to detect vacuum ultraviolet photons.
Charge signals induced on an array of 230,400 $7.5~\rm mm$ x
$8~\rm mm$ cathode pads are used to measure the position of the
Cherenkov photons using custom made low noise and high dynamic
range VA$\_$RICH ASIC's that convert the charge signals into
differential current signals in order to minimize cross talk in
the cables connecting the front end electronics to the data
boards. The back end electronics include transimpedance receivers
that transform the current signals into voltage signals which get
digitized afterwards by 12-bit differential ADC's in the data
boards located in the VME crates. More details can be found in
\cite{testbeam} and \cite{recogn}.

\section{PHYSICS PERFORMANCE}
\subsection{Photon Resolution and Photon Yield}
The physics performance studies were made previously
\cite{testbeam} using either Bhabha or hadronic
events\footnote{Here, photons that match the most likely mass
hypothesis within $\pm3\sigma$ were removed from consideration for
the other tracks to resolve overlaps between Cherenkov images for
different tracks.} where the single photon resolution parameter
$\sigma_{\theta}$ is the RMS width of the difference between the
measured and the expected single-photon Cherenkov angle
distribution. The photon yield per track is extracted from the
fitted and background subtracted photon yield per track
distribution. Consequently, the Cherenkov angle per track is found
as the arithmetic mean of all photoelectrons in an image within
$\pm 3 \sigma$ for each hypothesis.
A summary of the averaged values of these
parameters for flat and sawtooth radiators are shown in Table 1.

\begin{table*}[htb]
\vspace{0.4cm} \caption{The averaged values of the single-photon
resolution($\sigma_{\theta}$), the photon yield($N_{\gamma}$) and
the Cherenkov angle resolutions per track($\sigma_{track}$) from
Bhabha and hadronic CLEO III events, for flat and sawtooth
radiators.} \label{tab:Drecon} \vspace{0.2cm}
\begin{center}
\begin{tabular}{||c|c|c|c|c|}
\hline
Event Type & Type of Radiators & $\sigma_{\theta}$ $(\rm mrad)$  & ~$N_{\gamma}$~ & $\sigma_{track}$ $(\rm mrad)$\\
\hline
Bhabha & planar     & $14.7$        & $10.6$         & $4.7$ \\
       & sawtooth   & $12.2$        & $11.9$         & $3.6$ \\
\hline
Hadronic & planar     & $15.1$        &  $9.6$         & $4.9$ \\
         & sawtooth   & $13.2$        & $11.8$         & $3.7$ \\
\hline\hline
\end{tabular}
\end{center}
\vspace{0.3cm}
\end{table*}

\begin{figure} [htb]
\centerline{\hspace{.0in}\epsfxsize 2.6in
\epsffile{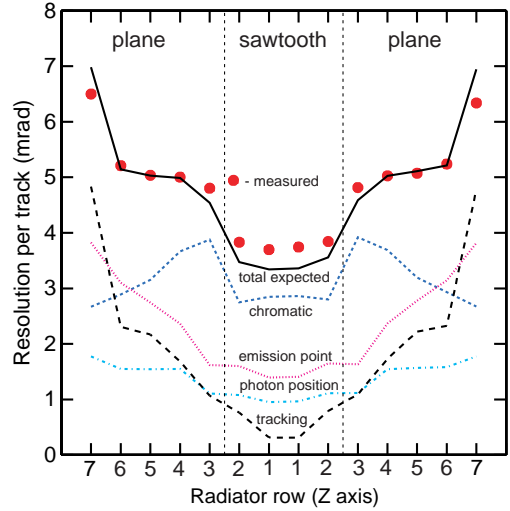}}
\caption{\label{res} Cherenkov angle resolution per
track versus radiator ring for Bhabha events from data (solid
points) and from the sum (solid line) of the different predicted
components (as labelled).} 
\end{figure}

The components of the Cherenkov angular resolution per track are
compared with the data as shown in Fig.~\ref{res}. The resolution
is mainly dominated by the chromatic dispersion and the error on
the photon emission point. Smaller components include the error on
the reconstructed photon position and the error on the charged
track's direction and position determination.

\subsection{Particle ID Likelihood: Definition and Operating Modes}

The particle identification criteria we are using for CLEO III and
CLEO-c analysis is different from what we presented in the last
section where only the optical path with the closest Cherenkov
angle to the expected one was considered. Here, the information on
the value of the Cherenkov angle and the photon yield for each
hypothesis is translated into a Likelihood of a given photon being
due to a particular particle. Contributions from all
photons\footnote{with a loose cut-off of $\pm5\sigma$.} associated
with a particular track are weighted by their optical
probabilities\footnote{ which include length of the radiation path
and the refraction probabilities.} then summed to form an overall
Likelihood denoted as $L_h$ for each particle hypothesis ``$h$``
($e$, $\mu$, $\pi$, $K$ or $p$), details about the analytical form
of the Likelihood function can be found in \cite{testbeam}.

\begin{figure} [htb]
\centerline{\epsfxsize
2.5in\epsffile{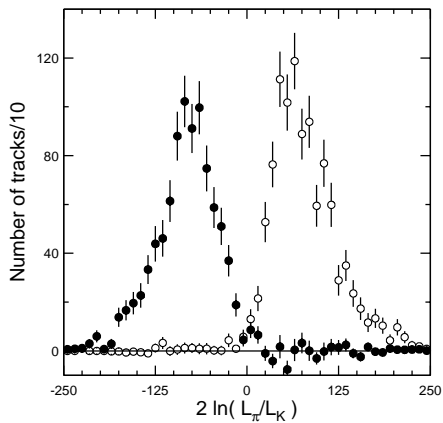}}
\caption{\label{likelihood_ratio}
      Distribution of $2\ln\left(L_\pi/L_K\right)$ $\sim\chi^2_K-\chi^2_\pi$
      for $1.0-1.5~\rm GeV/c$ kaons (filled)
      and pions (open) identified with the $D^*$ method.}
\end{figure}

The CLEO III data at $e^+e^-$ center-of-mass energies around
$10~\rm GeV$ have been used to evaluate the RICH performance.
Since the charge of the slow pion in the $D^{*+}\to\pi^+ D^0$
decay is opposite to the kaon charge in the subsequent $D^0\to
K^-\pi^+$ decay, the kaon and pion in the $D^0$ decay can be
identified without the RICH information. The efficiencies and fake
rates are hence extracted by studying the RICH identification
selectivity on the particle species selected with the $D^*$ tag.
Here the $D^0$ mass peak in the $K^-\pi^+$ mass distribution is
fitted to obtain the number of signal events for each momentum
interval. Fig.~\ref{likelihood_ratio} shows the distribution of
$2\ln\left(L_\pi/L_K\right)$, which is equivalent to the $\chi^2$
difference in the Gaussian approximation, for the identified kaons
and pions with $1.0-1.5~\rm GeV/c$ momentum.

\begin{figure} [hbt] \vspace{0.07cm}
\centerline{\epsfxsize
2.5in\epsffile{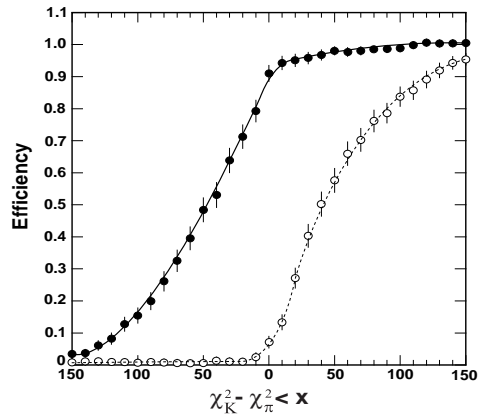}} 
\caption{\label{eff-fake}
      Kaon efficiency (filled points) and pion fake rate (hollow points) vs various cuts on the
      $\chi^2_K$-$\chi^2_\pi$ for tracks with $0.7-2.7~\rm GeV/c$ momentum.}
\end{figure}

\begin{figure} [hbt]
\vspace{0.4cm} \hspace{0cm}\centerline{\epsfxsize
2.4in\epsffile{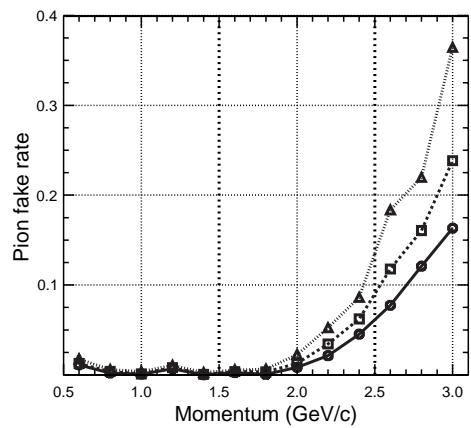}} 
\caption{\label{fake-vs-mom} Pion fake rate vs momentum for
different kaon efficiencies: 80\%\ (in circles), 85\%\ (in squares)
and 90\%\ (in triangles).} \vspace{0.3cm}
\end{figure}

\begin{figure} [htb]
\centerline{\epsfxsize 2.5in\epsffile{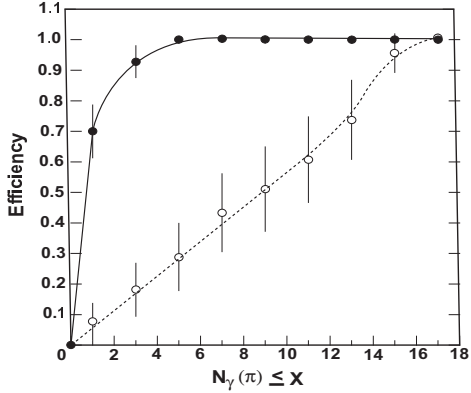}}
\caption{\label{rich-at-thresh} Kaon efficiency
(filled circles) and pion fake rate (empty circles) measured for
various cuts on the number of photons assigned to pion hypothesis
for tracks with $|p|<0.6~\rm GeV/c$.} 
\end{figure}

The detected fraction of kaons (pions) as a function of the cut on
$2\ln\left(L_\pi/L_K\right)$ is shown in Fig.~\ref{eff-fake}
and the pion fake rate for different kaon efficiencies versus
momentum is shown in Fig.~\ref{fake-vs-mom}.
Below $\thicksim$0.6$~\rm GeV$, the RICH can be used in the
threshold mode. Fig.~\ref{rich-at-thresh} shows the fraction of
kaons (pions) passing the cut restricting the number of photons
assigned to the pion hypothesis for tracks near and below the
Cherenkov radiation threshold for kaons ($0.44~\rm GeV/c$).

A summary of pion (kaon) efficiencies and kaon (pion) fake rates
from CLEO III data for $0.9-1.5~\rm GeV/c$ momentum range with a
$\chi^2_K-\chi^2_\pi$ cut at 0 is shown in Table 2.

\begin{table*}
\begin{center}
\caption{Particle Identification efficiencies ($\epsilon$) and
fake rates (F.R) from CLEO III and CLEO-c data pions and kaons
with momentum between $0.9$ and $1.5~\rm GeV/c$ and with a cut at
0 on the $\chi^2$ difference between kaons and pions from a
combined RICH and dE/dx information (the dE/dx doesn't have any
significant separation ability in this momentum range). Errors
here are statistical only.} \label{tab:Drecon}
\begin{tabular}{|c|c|c|c|c|c||}
\hline
Data type & Mom(GeV/c) &  $\epsilon_{\pi}$ $(\%)$  &  $K_{F.R}$ $(\%)$  &  $\epsilon_K$ $(\%)$  &  $\pi_{F.R}$ $(\%)$ \\
\hline
         & 0.9 &  $96.8 \pm 1.7$  &  $2.3 \pm 0.8$  &  $91.8 \pm 1.6$ &  $0.8 \pm 0.4$ \\
         & 1.1 &  $94.7 \pm 1.6$  &  $0.9 \pm 0.8$  &  $94.8 \pm 1.7$ &  $1.3 \pm 0.5$ \\
CLEO III & 1.3 &  $95.7 \pm 1.5$  &  $4.6 \pm 0.6$  &  $91.7 \pm 1.6$ &  $1.8 \pm 0.5$ \\
   & 1.5 &  $95.2 \pm 1.5$        &  $2.6 \pm 0.7$  &  $94.1 \pm 1.6$ &  $2.4 \pm 0.4$ \\
\hline 
CLEO-c   & 0.9 &  $95.1 \pm 2.1$  & $5.9 \pm 2.9$   & $87.2 \pm 3.6$  & $0.8 \pm 1.3$\\
 \hline \hline
\end{tabular}
\end{center}
\end{table*}

At lower center-of-mass energies near $4~\rm GeV$, the CLEO-c
program has started extensive studies of charm meson decays. In
these analyses, one of the $D$'s is reconstructed through hadronic
channels while the other $\overline{D}$ is used as a signal side
for various studies. For $\pi/K$ ID efficiency measurements pions
from $D^0\to K\pi\pi^0$, $D^0\to K_s\pi\pi$ and $D^+ \to
K^-\pi^+\pi^+$ decays and kaons from $D^0\to K\pi\pi^0$ and $D^+
\to K^-\pi^+\pi^+$ decays are used. The particle identification
efficiency is defined in this case as the ratio of the number of
$D^0$ events that passed the particle ID criteria to the number of
$D^0$ events without any PID. Efficiencies and fake rates for
$0.9~\rm GeV/c$ momentum pions and kaons from the data with a
$\chi^2_K-\chi^2_\pi$ cut at 0 are summarized at the end of Table
2.

\subsection{Example of Particle ID Performance}

\begin{figure} [hbt]
\centerline{\epsfxsize
2.45in\epsffile{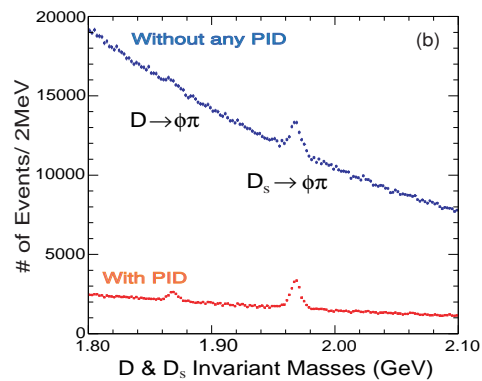}}
\caption{\label{ds} The invariant mass of $D$ and $D_s$ candidates
from the $\Upsilon$(4S) resonance data collected with the CLEO III
detector without any PID (blue curve) and with RICH and dE/dx PID
applied on just one of the
kaons 
(red curve).}
\end{figure}

We used recently $0.42~\rm fb^{-1}$ of data taken on the
$\Upsilon$(5S) resonance, $6.34~\rm fb^{-1}$ of data collected on
the $\Upsilon$(4S) and $2.32~\rm fb^{-1}$ of data taken in the
continuum below the $\Upsilon$(4S) with the CLEO III detector to
measure the branching fraction B$(\Upsilon$(5S)$\to B_s^{(*)}
\overline{B_s^{(*)}} )$ \cite{5S} which has never been measured
before. In this analysis, we reconstructed $D_s$ mesons through
the decay mode: $D_{s} \to \phi\pi$ and $\phi \to KK$ where we
used the RICH information to identify one of the kaons with a
momentum higher than $0.62~\rm GeV/c$. We show in Fig.~\ref{ds}
for instance the large combinatoric backgrounds, from the
$\Upsilon$(4S) on resonance data mentioned above, that we would
have included in the $D_s$ candidates invariant mass spectrum if
we didn't take advantage of the particle identification of one of
the kaons.

\section{Conclusions}
The CLEO LiF-TEA RICH is providing us with excellent particle
identification for all momenta relevant to the CLEO III beauty
threshold data and present charm threshold CLEO–c data. It has
operated successfully for over 4 years.

We have made and are making extensive studies of the Upsilon, $B$
and $B_s$ decays and, since last year, we have used the detector
for the CLEO-c program to study charm mesons and charmonium
decays. Thus, the physics performance of the CLEO RICH detector
has met the benchmarks and the design criteria.

\section{Acknowledgments}
I gratefully acknowledge professors: S. Stone, M. Artuso, and T.
Skwarnicki for the valuable discussions and comments and my CLEO
collaborators especially G. Tatishvili for the valuable input and
N. Menaa for the stimulating discussions. I would like also to
thank the CESR staff for the excellent luminosity and the National
Science Foundation for supporting this work.

\end{document}